\documentclass[proceedings]{JHEP3}
\usepackage{epsfig,multicol,axodraw,color}
\newbox\mybox
\newcommand\fverb{\setbox\mybox=\hbox\bgroup\verb}
\newcommand\fverbdo{\egroup\medskip\noindent\fbox{\unhbox\mybox}\ }
\newcommand\fverbit{\egroup\item[\fbox{\unhbox\mybox}]}

\conference{International Workshop on Astroparticle and High Energy Physics}

\newcommand{\etal} {{\it et al.\/}}

\title{Degenerate neutrinos from a supersymmetric $A_4$ model}

\author{M. Hirsch${}^1$,  J. C. Rom\~ao${}^{2}$, 
\speaker{S. Skadhauge}${}^{2}$,
J. W. F. Valle${}^1$, A. Villanova del Moral${}^1$ \\ 
{\it $^1$ AHEP Group, Instituto de F\'{\i}sica Corpuscular --
  C.S.I.C./Universitat de Val{\`e}ncia \\
  Edificio de Institutos de Paterna, Apartado 22085,
  E--46071 Val{\`e}ncia, Spain\\}
{\it $^2$ Departamento de F\'\i sica and CFIF, Instituto Superior T\'ecnico\\
          Av. Rovisco Pais 1, $\:\:$ 1049-001 Lisboa, Portugal\\}}

\abstract{
We investigate the supersymmetric $A_4$ model recently 
proposed by Babu, Ma and Valle \cite{bmv}. 
The model naturally gives quasi-degenerate neutrinos that are  
bi-largely mixed, in agreement with observations. 
Furthermore, the mixings in the quark sector are constrained to be small, 
making it a complete model of the flavor structure. Moreover, it  
has the interesting property that CP-violation in the leptonic 
sector is maximal (unless vanishing). 
The model exhibit a close relation between the slepton 
and lepton sectors and we derive the slepton spectra 
that are compatible with neutrino data and the present 
bounds on flavor-violating charged lepton decays. 
The prediction for the branching ratio of the 
decay $\tau \rightarrow \mu \gamma$ has a lower limit of $10^{-9}$. 
In addition, the overall neutrino mass scale is constrained to 
be larger than 0.3 eV. 
Thus, the model will be tested in the very near future.}

\begin{document} 
\section{Introduction}
 
Thanks to the tremendous experimental progress in neutrino 
physics~\cite{skatm,sksolar,sno,k2k,kamland,chooz} the main 
features of the mixing matrix in the leptonic sector 
have now been established. Especially, it is evident that the 
leptonic and the quark mixing matrices are rather different. 
The rough structure of the mixings being two large and one small 
mixing angle in the leptonic sector as opposed to three small 
mixings in the quarks sector. This calls for further studies 
of possible origins of the flavor structure. 

Building attractive models which incorporates the structure of both the 
quark and lepton mixing matrices is a major challenge. 
In this work~\cite{us} we will examine a model that achieves this in 
a rather simple way. The main ingredients in the model, which was 
put forward in Ref.\cite{bmv}, are;
(I) $A_4$ family symmetry, where $A_4$ is the symmetry group of the 
tetrahedron. (II) Heavy singlet lepton, quark and Higgs fields.  
(III) Low energy supersymmetry. 
(IV) Flavor violating soft SUSY breaking terms.  
The philosophy is to generate the mixing matrices through 
radiative corrections, starting from different boundary condition in the 
quark and the leptonic sector.  
These boundary conditions are provided by the family symmetry, which 
fix the value of the mixing matrices at some high energy scale, $M_N$, 
where the family symmetry is spontaneously broken.
The quark mass matrices are aligned at $M_N$ and therefore the mixing 
angles vanish at the high energy scale. 
Due to the hierarchical structure of the quark mass, 
the small radiative corrections naturally give small mixing angles. 
The observed low energy $V_{CKM}$ can indeed be achieved, starting from 
the identity matrix at some high energy scale~\cite{radVckm}. 
The neutrino masses are on the other hand fixed to be degenerate at $M_N$, 
thereby large mixings are easily obtained through quantum corrections.
In this way the solar angle is naturally large, whereas the two other 
angles are restricted by the family symmetry. Also, the model has the 
interesting qualities of maximal CP-violation (unless zero) 
in the leptonic sector and exact atmospheric mixing.  

Let us look at the sources of the radiative corrections to the 
mass matrices. Using the standard MSSM renormalization group equations 
(RGE's) and degenerate neutrinos at some high energy scale, it is not 
possible to obtain a suitable neutrino spectrum~\cite{nogodegen}. 
Nonetheless, allowing for threshold corrections from 
non-standard flavor violating (FV) interactions this is no 
longer true \cite{sleptonFV}. 
We explicitly show that including the radiative corrections coming 
from FV scalar lepton (slepton) interactions, a realistic low energy 
neutrino mass matrix can be obtained. 
This implies that both the atmospheric and the solar 
mass squared differences are generated by the radiative corrections. 

A crucial ingredient in the model is therefore the flavor-non-blindness of 
the soft SUSY breaking terms, thus potentially large SUSY contribution to 
lepton flavor violation (LFV) may arise.  
However, the experimental bounds on FV in the lepton sector are 
less severe than in the quark sector. Especially, the bounds on the 
second and third generations are quite weak.   
The mechanism responsible for splitting the neutrino masses 
and the bounds on LFV push the off-diagonal elements of the slepton mass 
matrix in different directions. 

We derive the possible low energy slepton masses and mixings by 
using the present knowledge of the neutrino mass matrix and the 
constraints from lepton flavor violation. 
We show that the model is indeed viable, although it is severely 
constrained. We give the predictions for lepton flavor violations 
processes, such as $\tau \rightarrow \mu \gamma$ and the overall 
neutrino mass scale. 
These are within experimental reach in the very near future.  
 
\section{Outline of the model}
We present a schematic overview of the model, for further detail we 
refer to Refs.\cite{bmv,us,a4}. Let us briefly discuss the $A_4$ group 
as it is not widely known.
It is the symmetry group of even permutations of four 
elements and has four irreducible representation; 
three independent singlet, which we denote as 
$\underline{1},\underline{1}' $ and $\underline{1}''$ and one triplet 
$\underline{3}$.  
The decomposition of the product of two triplets is
\begin{equation}
  \underline{3}\times \underline{3} = \underline{1}+
\underline{1'}+ \underline{1''}+\underline{3}+ \underline{3} \, .
\end{equation}
Defining $\underline{3}_i=(a_i,b_i,c_i),i=1,2$ we have e.g. the 
following rules
\begin{eqnarray}\label{decomp}
\underline{1} & =&  a_1 a_2 + b_1 b_2+ c_1c_2  \,,\nonumber \\
\underline{1'}&=&a_1 a_2 + \omega^2 b_1b_2+ \omega c_1c_2 \, , \\
\underline{1''}&=&a_1 a_2 + \omega b_1b_2+ \omega^2 c_1c_2 \, .\nonumber
\end{eqnarray}

The usual MSSM fields are assigned the following transformation properties 
under $A_4$,
\begin{equation}
 \hat{Q}_i = (\hat{u}_i,\hat{d}_i), \,\,\, \hat{L}_i=(\hat{\nu}_i,\hat{e}_i) 
\sim \underline{3}, \,\,\,\, \,\,\hat{\phi}_{1,2}
\sim \underline{1}
\end{equation}
\begin{equation}
\hat{u}_1^c,\hat{d}_1^c,\hat{e}_1^c \sim \underline{1},
\,\,\,\,\,\, 
\hat{u}_2^c,\hat{d}_2^c,\hat{e}_2^c \sim \underline{1}',
\,\,\,\,\,\, 
\hat{u}_3^c,\hat{d}_3^c,\hat{e}_3^c \sim \underline{1}''  \, .
\end{equation}
We add heavy quark, lepton and Higgs superfields
as follows   
\begin{equation}
 \hat{U}_i, \,\, \hat{U}_i^c, \,\, \hat{D}_i, \,\, \hat{D}_i^c,  \,\,
 \hat{E}_i, \,\, \hat{E}_i^c, \,\, \hat{N}_i^c, \,\, 
 \hat{\chi}_i, \sim \underline{3}  
\end{equation}
all are $SU(2)$ singlets.
The superpotential is then given by
\begin{eqnarray}\label{superW}
 \hat{W} &=& M_U \hat{U}_i \hat{U}_i^c+f_u \hat{Q}_i\hat{U}_i^c\hat{\phi}_2  
 + h_{ijk}^u \hat{U}_i \hat{u}_j^c \hat{\chi}_k \nonumber \\
         &+& M_D \hat{D}_i \hat{D}_i^c+f_d \hat{Q}_i\hat{D}_i^c \hat{\phi}_1  
 + h_{ijk}^d \hat{D}_i \hat{d}_j^c \hat{\chi}_k \nonumber \\
         &+& M_E \hat{E}_i \hat{E}_i^c+f_e \hat{L}_i\hat{E}_i^c \hat{\phi}_1  
 + h_{ijk}^e \hat{E}_i \hat{e}_j^c \hat{\chi}_k  \\
         &+& \frac{1}{2}M_N \hat{N}_i^c\hat{N}_i^c + 
             f_N\hat{L}_i\hat{N}_i^c\hat{\phi}_2  + 
             \mu \hat{\phi}_1\hat{\phi}_2  \nonumber  \\
         &+& \frac{1}{2}M_{\chi}\hat{\chi}_i \hat{\chi}_i 
+ h_{\chi} \hat{\chi}_1\hat{\chi}_2\hat{\chi}_3 \nonumber \, ,
\end{eqnarray}
where an extra $Z_3$ symmetry, softly broken by $M_\chi$, has been 
implemented in order to restrict further the superpotential. 
The $A_4$ symmetry is spontaneously broken at the high scale by the 
vev's $\langle \chi_i \rangle = u =-M_\chi/h_\chi, i=1,2,3$. 
This is a F-flat direction and therefore it preserves 
SUSY, which is only broken at the TeV energy scale.   
The low energy effective theory of the model is in fact the MSSM. 
The electroweak symmetry is broken by the vev's, 
$\langle \phi_i^0 \rangle = v_i$ of the two Higgs doublet. 
We define as usual $\tan(\beta)=v_2/v_1$.

The charged lepton mass matrix is given by the simple form 
\begin{equation}\label{clepmass}
M_{eE}= \left(\matrix{ 
 0 & 0 & 0 & f_e v_1 & 0 & 0 \cr
 0 & 0 & 0 & 0 & f_e v_1 & 0 \cr
 0 & 0 & 0 & 0 & 0 & f_e v_1 \cr
 h_1^e u & h_2^e u & h_3^e u & M_E & 0 & 0 \cr
 h_1^e u & h_2^e \omega u 
 & h_3^e \omega^2 u & 0 & M_E & 0 \cr
 h_1^e u & h_2^e \omega^2 u 
 & h_3^e \omega u & 0 & 0 & M_E \cr} \right) \;, 
\end{equation}
in the basis $(e_i,E_i)$. Note that the factors of $\omega$ arise from 
forming $A_4$ invariant products (see Eq.\ref{decomp}).  
This mass matrix is of see-saw type, and the 
effective $3 \times 3$ low energy mass matrix, 
$M_{\ell}^{\rm eff}$, can be diagonalized by 
$M_{\ell}^{\rm diag.}=U_L M_{\ell}^{\rm eff} I$, where 
the left diagonalization matrix reads
\begin{equation}
U_{L}=\frac{1}{\sqrt{3}}\left(\matrix{ 1 & 1 & 1 \cr
1 & \omega & \omega^2 \cr
1 & \omega^2 & \omega \cr} \right)  \;.
\end{equation}
The Yukawa couplings $h_i^e$, $i=1,2,3$, are chosen such that 
the three eigenvalues of $M_{\ell}^{\rm eff}$, 
agree with the measured masses of the electron, muon and tau particles. 
As also the up-type and down-type quark mass matrices have the same 
structure as in Eq.(\ref{clepmass}), they will be simultaneously 
diagonalized, enforcing $V_{\rm CKM} = I$ at the high scale. 
The observed low-energy $V_{\rm CKM}$ can be obtained from non-standard 
radiative corrections~\cite{radVckm}. 
In the numerical study, however, we only consider the 
leptonic sector. 

The Majorana neutrino mass matrix in the flavor basis 
$(\nu_e,\nu_\mu,\nu_\tau,N_1^c,N_2^c,N_3^c)$,  may be written as
\begin{equation}
\left(\matrix{ 0 & U_L f_N v_2 \cr
U_L^T f_N v_2 & M_N \cr} \right) \;, 
\end{equation}
where $M_N$ is proportional to the identity matrix. 
This is the usual see-saw form, and the effective low-energy 
neutrino mass matrix, is given by
\begin{equation}\label{mnutree}
 M_\nu^0 = \frac{f_N^2 v_2^2}{M_N}U_L^TU_L= \frac{f_N^2 v_2^2}{M_N}
\lambda_0 \,,\qquad
\lambda_0= 
\left(\matrix{ 1 & 0 & 0 \cr 0 & 0 & 1 \cr 0 & 1 & 0 \cr} \right)
\;.
\end{equation}
Therefore, a very attractive tree-level neutrino mass matrix has been 
obtained. It has degenerate neutrinos, $m_1=m_2=m_3$, 
and exact maximal atmospheric mixing. 
Also the tree-level value of the 'reactor' angle, $s_{13}$, is 
vanishing\footnote{We define the most split neutrino mass as $m_3$ 
and require $m_2>m_1$, therefore the 'reactor' angle is always given 
by $s_{13}$}. Here, we use the standard parametrization of the 
neutrino mixing matrix as in Ref.\cite{pgd}.

\FIGURE{
\begin{picture}(420,130)(0,-50)
\ArrowLine(0,10)(45,10)
\ArrowLine(45,10)(120,10)
\ArrowLine(120,10)(155,10)
\ArrowLine(200,10)(155,10)
\DashCArc(80,10)(40,0,180){3}
\Vertex(155,10){3} 
\DashLine(155,10)(130,-30){3}
\DashLine(155,10)(180,-30){3}
\Text(10,20)[]{$\nu_{j}$}
\Text(80,65)[]{$\tilde\ell/ \tilde\nu$}
\Text(135,20)[]{$\nu_{k}$}
\Text(80,-3)[]{$\chi^{+}/\chi^0$}
\Text(185,20)[]{$\nu_{i}^c$}
\Text(10,70)[]{\Large{(a)}}
\ArrowLine(220,10)(265,10)
\ArrowLine(300,10)(265,10)
\ArrowLine(375,10)(300,10)
\ArrowLine(420,10)(375,10)
\DashCArc(340,10)(40,0,180){3}
\Vertex(265,10){3} 
\DashLine(265,10)(240,-30){3}
\DashLine(265,10)(290,-30){3}
\Text(230,20)[]{$\nu_{j}$}
\Text(340,65)[]{$\tilde\ell/\tilde\nu$}
\Text(285,20)[]{$\nu_{k}^c$}
\Text(340,-3)[]{$\chi^{+}/\chi^0$}
\Text(405,20)[]{$\nu_{i}^c$}
\Text(230,70)[]{\Large{(b)}}
\end{picture}
\begin{picture}(420,150)(0,-50)
\ArrowLine(0,10)(45,10)
\ArrowLine(45,10)(100,10)
\ArrowLine(100,10)(155,10)
\ArrowLine(200,10)(155,10)
\DashCArc(100,10)(55,0,180){3}
\Vertex(155,10){3}
\DashLine(155,10)(155,-30){3}
\DashLine(100,10)(100,-30){3}
\Text(10,20)[]{$\nu_{j}$}
\Text(100,85)[]{$\tilde\ell/\tilde\nu$}
\Text(70,-3)[]{$\chi^{+}/\chi^0$}
\Text(130,-3)[]{$\chi^{+}/\chi^0$}
\Text(185,20)[]{$\nu_{i}^c$}
\Text(10,90)[]{\Large{(c)}}
\ArrowLine(220,10)(265,10)
\ArrowLine(320,10)(265,10)
\ArrowLine(375,10)(320,10)
\ArrowLine(420,10)(375,10)
\DashCArc(320,10)(55,0,180){3}
\Vertex(265,10){3}
\DashLine(265,10)(265,-30){3}
\DashLine(320,10)(320,-30){3}
\Text(230,20)[]{$\nu_{j}$}
\Text(320,85)[]{$\tilde\ell/\tilde\nu$}
\Text(290,-3)[]{$\chi^+/\chi^0$}
\Text(350,-3)[]{$\chi^+/\chi^0$}
\Text(405,20)[]{$\nu_{i}^c$}
\Text(230,90)[]{\Large{(d)}}
\end{picture}
\caption{Feynman diagrams responsible for the radiative 
corrections to mass of the neutrino. 
The fat vertex indicates an effective Lagrangian term obtained 
from integrating out the heavy right-handed neutrinos.}
\label{loopfig}
}
Consider now the quantum corrections to the tree-level matrix in 
Eq.(\ref{mnutree}) written at the scale $M_N$.  
In general there are two kinds of corrections; 
the standard MSSM renormalization group equation (RGE) effects 
and the low-energy threshold corrections. 
The RGE effects, accounting for the running from $M_N$ to 
the SUSY scale $M_S$, can in the flavor basis be approximated 
by~\cite{renormneu}
\begin{equation}
  M_{\alpha\beta}(M_S) \simeq \left[1-\frac{m_{\alpha}^2+m_{\beta}^2}
{16\pi^2 v^2\cos^2(\beta)}\log(M_N/M_S)\right] M_{\alpha\beta}(M_N) 
 \;.
\end{equation} 
These effects can not produce corrections to the textures zeros in 
$M_\nu^0$, as they are proportional to the original mass element.
Nonetheless, it is clear that small corrections to the tree-level 
texture zeros are necessary in order to obtain a realistic mass matrix 
and that these must originate from flavor-changing interactions. 
Therefore, we invoke flavor-changing soft SUSY breaking terms and 
indeed in this case the threshold corrections can give the desired effect.
In fact, in SUSY theories the threshold corrections can dominate 
over the RGE effects.  

The one-loop (threshold) contributions to the neutrino 
mass renormalization are given by the diagrams shown in Fig.\ref{loopfig}. 
For the evaluation we make the following approximations:  
\begin{itemize}
\item We consider only the $3 \times 3$ left-left, $M_{LL,\ell}$, 
part of the charged slepton mass matrix. 
\item We assume $M_{LL,\tilde\ell}^2=M_{LL,\tilde\nu}^2=M_{LL}^2$, thereby 
neglecting the, usually much smaller, D-term contributions. 
\end{itemize}
Consequently, the sleptons and sneutrinos have identical masses and mixings.
The contribution from the $\chi^+/\tilde{\ell}_R$ loop in the 
diagrams in Fig.\ref{loopfig} are suppressed with the Yukawa couplings 
squared. Hence, for the first two generations and also the third generation 
for small $\tan(\beta)$ the above approximation 
is reasonable. As we will discuss below, the viable solutions 
have indeed small $\tan(\beta)$.
It is easily realized that the one-loop corrections to the 
neutrino mass matrix have the structure
\begin{equation}
\lambda^{\rm 1-loop} = \lambda^0 \hat{\delta} + (\hat{\delta})^T \lambda^0 
\;, \qquad \hat{\delta}=
\left(\matrix{ \delta_{ee} & \delta_{e\mu}^* & \delta_{e\tau}^* \cr
\delta_{e\mu} & \delta_{\mu\mu} & \delta_{\mu\tau}^* \cr
\delta_{e\tau} & \delta_{\mu\tau} & \delta_{\tau\tau} \cr}\right) \,.
\end{equation}
The analytic expression for $\hat{\delta}$ may be found in 
Ref.\cite{us}. 
The leading effect from the RGE running, originating from the 
tau Yukawa coupling, is taken into account. 
However, the arguments below are 
valid considering the full RGE contribution. 
The form of the neutrino mass matrix may be written as   
\begin{equation}\label{neumass}
M_{\nu}^{\rm 1-loop}= m_0
\left(\matrix{1+\delta_0+2\delta+2 \delta' & \delta'' & \delta''^* \cr 
 \delta'' & \delta & 1 + \delta_0+ \delta - 2\delta_\tau\cr
\delta''^* & 1+\delta_0+\delta -2\delta_\tau & \delta \cr}\right) \;,
\end{equation}
where $m_0$ is the overall neutrino mass scale. All parameters expect 
$\delta''$ may be taken real\footnote{The phase of the $\delta$ 
($\delta_{\mu\tau}$) 
parameter can be rotated away, even though the neutrinos are Majorana 
particles, due to the special form of $M_{\nu}^{\rm 1-loop}$.}.  
Here, we have defined
\begin{equation}\label{raddelta}
  \delta = 2 \delta_{\mu\tau} \;, \qquad \delta_0= \delta_{\mu\mu}+
\delta_{\tau\tau}-2\delta_{\mu\tau}
\end{equation}
\begin{equation}\label{raddeltap}
  \delta '=\delta_{ee} -\delta_{\mu\mu}/2 -\delta_{\tau\tau}/2-\delta_{\mu\tau}
\end{equation}
\begin{equation}\label{raddeltapp}
  \delta ''= \delta_{e\mu}^* + \delta_{e\tau} 
\end{equation}
\begin{equation}
  \delta_\tau \equiv
\frac{m_{\tau}^2}{8\pi^2 v^2\cos^2(\beta)}\log(M_N/M_S)
\end{equation}  
The effect of $\delta_0$ can be absorbed into $m_0$, as its value 
does not affect the mixing angles. 
In the numerical analysis we assume that $\delta''$ is real, 
in which case the neutrino mass matrix can be diagonalized analytically.
The eigenvalues are
\begin {eqnarray}
m_1 &=& 
m_0\left( 1+2\delta+\delta'-\delta_\tau 
-\sqrt{\delta'^2+2\delta''^2+2\delta'\delta_\tau+\delta_\tau^2}
\right) \nonumber \\
m_2 &=& 
m_0\left(1+2\delta+\delta'-\delta_\tau 
+\sqrt{\delta'^2+2\delta''^2+2\delta'\delta_\tau+\delta_\tau^2}
\right) \\
m_3 &=& m_0(-1+2\delta_\tau) \nonumber
\end{eqnarray}
Hence, the mass squared differences can be approximated by
\begin{equation}\label{dmsqatm} 
 \Delta m^2_{\rm atm} \simeq 4 m_0^2 \delta
\end{equation}
\begin{equation}\label{dmsqsol}
 \Delta m^2_{\rm sol} \simeq 4 m_0^2  
 \sqrt{\delta'^2+2\delta''^2+2\delta'\delta_\tau+\delta_\tau^2}
\end{equation}
Although the neutrinos are quasi-degenerate, we will refer to the case 
of $\delta<0$ as normal hierarchy and of $\delta>0$ as inverted hierarchy.
The solar angle is given by
\begin{equation}\label{solangle}
   \tan^2(\theta_{\rm sol})=
\frac{2\delta''^2}{(\delta'+ \delta_\tau -
 \sqrt{\delta'^2+2\delta''^2+2\delta'\delta_\tau+\delta_\tau^2})^2}
\end{equation} 
and is independent of both $\delta$ and $\delta_0$.

The form of $M_{\nu}$ in Eq.(\ref{neumass}) implies that the mixing matrix 
satisfies $|U_{\mu i}| =|U_{\tau i}|, \,\, i=1,2,3$ \cite{maxcp}. 
From this it is easily seen that the atmospheric mixing angle is 
{\it maximal}, and thus not affected by the radiative corrections~\cite{bmv}.  
The relation also implies that $s_{13} \cos(\delta_{\rm CP}) = 0$, 
where $\delta_{\rm CP}$ is the Dirac CP-phase. Therefore, unavoidably 
along with a non-zero value of $s_{13}$ is  maximal CP-violation in 
the leptonic sector~\cite{bmv}. 
The Majorana phases are constrained to be 1 or $i$, 
and will not give rise to CP-violation~\cite{maxcp}.
The property of maximal CP-violation gives interesting perspectives for 
discovery of leptonic CP-violation in future long-baseline experiments. 

In Eq.(\ref{neumass}) the neutrino mass matrix is written in the 
tree-level flavor basis. 
The radiative corrections to the charged lepton mass matrix  
will therefore result in small corrections to the leptonic mixing 
angles. These corrections, albeit very small, will 
result in deviation from maximal atmospheric mixing and also 
maximal CP-violation. 

\section{Analysis}
\FIGURE[t]{
\includegraphics[width=0.45\textwidth]{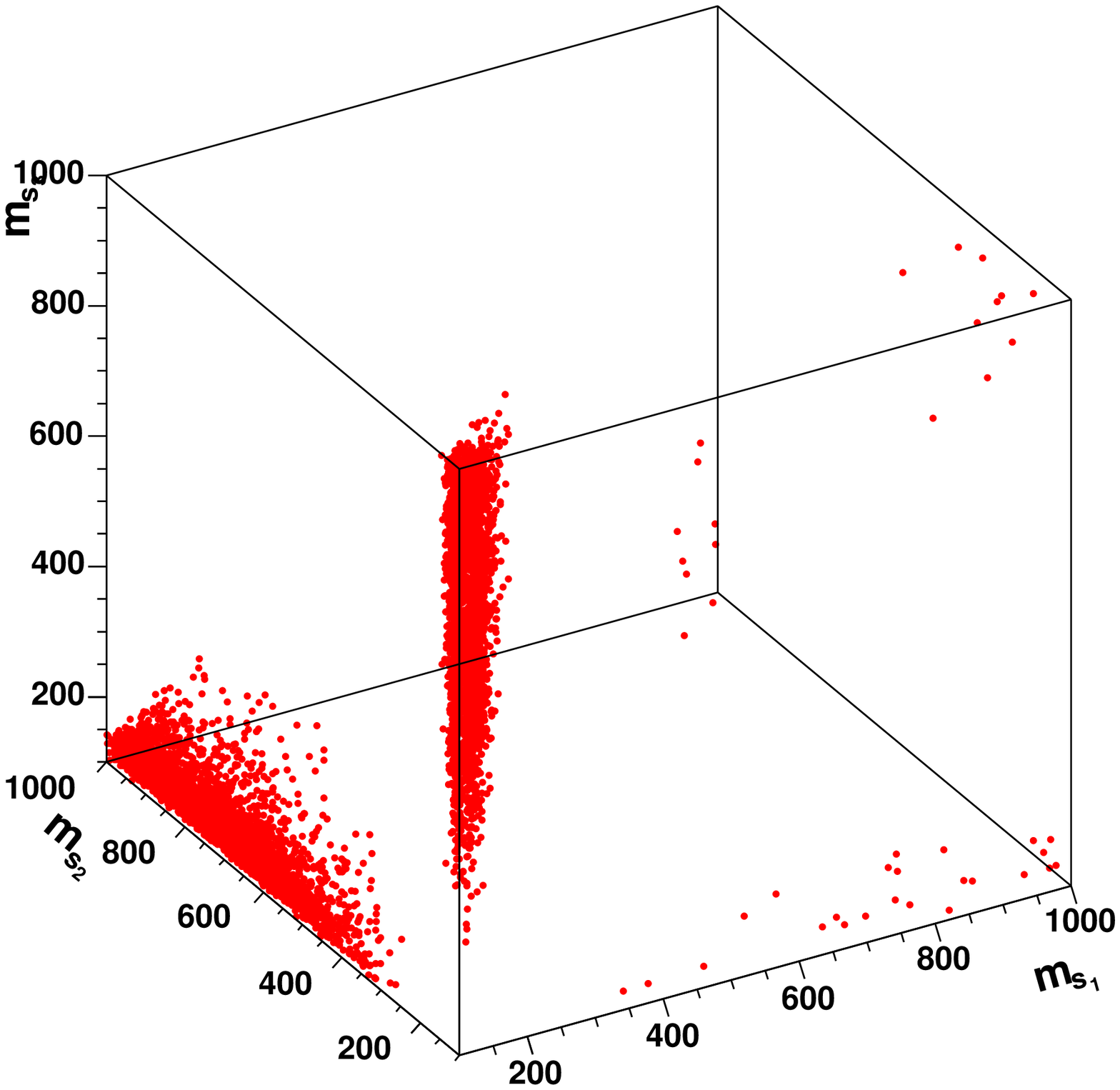}
\includegraphics[width=0.45\textwidth]{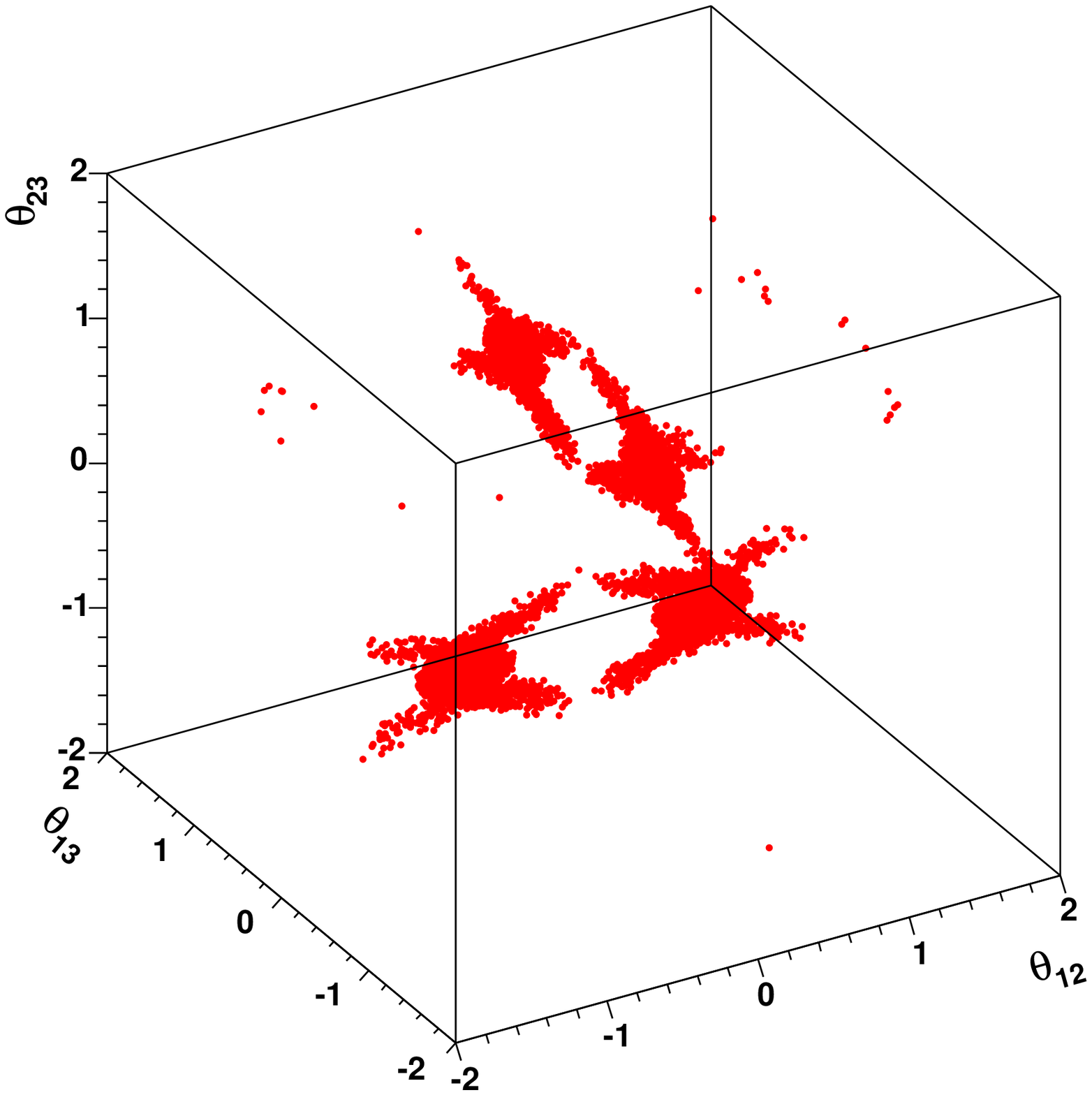}
 \caption{The allowed range for the slepton masses and mixings for $\delta<0$.}
\label{massfig}
}
\FIGURE[t]{
\begin{picture}(360,180)(0,0)
\CBox(0,160)(1,175){Black}{Blue}
\CBox(1,160)(80,175){Black}{Yellow}
\CBox(80,160)(160,175){Black}{Red}
\CBox(0,25)(64,40){Black}{Blue}
\CBox(64,25)(112,40){Black}{Yellow}
\CBox(112,25)(160,40){Black}{Red}
\CBox(0,5)(96,20){Black}{Blue}
\Text(48,12)[c]{\textcolor{white}{$\tilde{e}$}}
\CBox(96,5)(128,20){Black}{Yellow}
\Text(112,12)[c]{\textcolor{black}{$\tilde{\mu}$}}
\CBox(128,5)(160,20){Black}{Red}
\Text(144,12)[c]{\textcolor{black}{$\tilde{\tau}$}}
\CBox(200,5)(201,20){Black}{Blue}
\CBox(201,5)(280,20){Black}{Yellow}
\CBox(280,5)(360,20){Black}{Red}
\CBox(200,160)(296,175){Black}{Blue}
\Text(248,167)[c]{\textcolor{white}{$\tilde{e}$}}
\CBox(296,160)(328,175){Black}{Yellow}
\Text(312,167)[c]{\textcolor{black}{$\tilde{\mu}$}}
\CBox(328,160)(360,175){Black}{Red}
\Text(345,167)[c]{\textcolor{black}{$\tilde{\tau}$}}
\CBox(200,140)(264,155){Black}{Blue}
\CBox(264,140)(312,155){Black}{Yellow}
\CBox(312,140)(360,155){Black}{Red}
\end{picture}
\caption{The rough form of the slepton and sneutrino spectrum 
in the case of normal hierarchy (left) or inverted hierarchy (right).
The dark shaded area represent the amount of selectron in 
the slepton mass-eigenstate ($U_{ej}^2$), the light shaded 
the amount of smuon and finally the medium shaded the amount of stau.}
\label{snuspecfig}
}

We carry out a detailed numerical analysis, using 
constraints from neutrino data and LFV. 
The allowed parameter space is determined by 
a random search through the 10 dimensional parameter space, 
keeping all SUSY masses real and in the range 100 GeV to 1000 GeV. 
For the neutrino parameters we use the 3$\sigma$ 
allowed ranges~\cite{Maltoni}.
The strongest bounds on LFV comes from $\ell_j \rightarrow \ell_i \gamma$ 
and explicitly formulas for the SUSY contributions can be found in 
Ref.\cite{lfv}. We have compared our numerical programs for the 
branching ratios against the ones in Ref.\cite{mario} and 
found agreement. The present bounds are~\cite{pgd}  
\begin{eqnarray}
 BR(\mu \rightarrow e \gamma) &<& 1.2 \times 10^{-11} \;,\nonumber \\
 BR(\tau \rightarrow \mu \gamma) &<& 1.1 \times 10^{-6} \;, \\
 BR(\tau \rightarrow e \gamma) &<& 2.7 \times 10^{-6} \nonumber \;.
\end{eqnarray}
The overall neutrino mass scale $m_0$ for the degenerate neutrinos is 
constrained by $(\beta\beta)_{0\nu}$ experiments and cosmology. 
The bound from neutrinoless double beta decay is less restrictive considering 
the uncertainties in the nuclear matrix element. 
Here we will apply the cosmological bound of 0.6 eV 
derived in Ref.\cite{tegmark}. 

It turns out difficult to get the neutrino mass splittings large enough.
The mass splittings are related to the parameters $\delta,\delta'$ 
and $\delta''$ in Eqs. \ref{raddelta}, \ref{raddeltap}, \ref{raddeltapp}, 
and these are increasing functions of the slepton mixings and 
also mass splittings. This is potentially in conflict with the 
restrictions on LFV which will push the spectrum toward mass degeneracy 
and small mixings. Therefore it is a non-trivial task to show that a viable 
spectrum exist. 

\FIGURE{
  \includegraphics[width=0.55\textwidth]{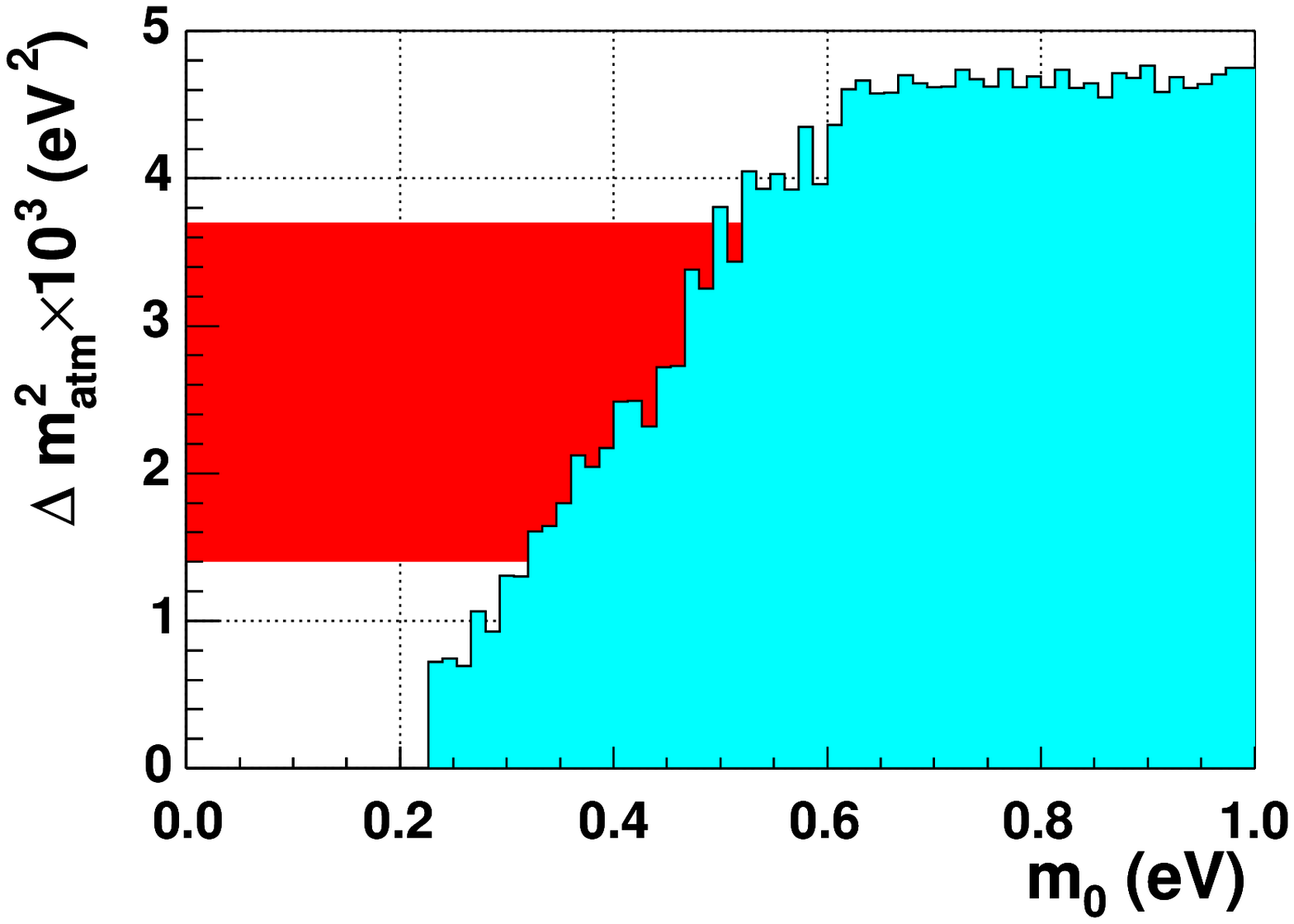}
\caption{The light shaded histogram shows the maximum possible value 
of the atmospheric mass squared difference as a function of $m_0$. 
The dark shaded region is the current 3$\sigma$ allowed region for 
$\Delta m^2_{\rm atm}$.}
\label{mzerofig}
}
Indeed, the allowed range for the charged slepton parameters, 
identical to that of the sneutrinos, 
is quite restricted. The spectra fall into two different groups. 
The normal hierarchy having two low mass sleptons ($\sim 150$ GeV) 
and one heavy (above $\sim$ 500 GeV), and 
the inverted hierarchy case having two heavy sleptons and 
one ligth. In both cases at least one slepton mass is below 
about 200 GeV, which is detectable at the LHC. 
Most points fall into the case of normal hierarchy, which often   
corresponds to a normal hierarchy for the neutrinos as well. 
In Fig.\ref{massfig} the slepton masses and mixings are shown 
for the case of $\delta<0$. 
The typical case has one small and two large mixing angles. 
This structure is also found for $\delta>0$. 
Evidently the small mixing angle is needed to 
suppress the decay $\mu \rightarrow e \gamma$. Also the degeneracy of 
two of the sleptons helps to minimize the LFV. As a rule of thumb 
there is at least one pair of sleptons with a mass splitting of less than 
40 GeV. The rough spectrum needed is schematized in Fig.\ref{snuspecfig},  
although there is room for substantial deviations from the shown spectrum. 
The similarity with the neutrino spectrum is quite striking and is related 
to the necessary relation $\delta \gg \delta', \delta''$.  

The parameter $\delta$ is bounded from above which, through the 
relation to $\Delta m^2_{\rm atm}$ in Eq.\ref{dmsqatm}, 
imposes a minimum value for $m_0$. In Fig.\ref{mzerofig} 
the maximum achievable value of $\Delta m_{\rm atm}^2$ is plotted 
against the value of $m_0$. The lower bound   
\begin{equation}
 m_0 > 0.3 \, {\rm eV} \;
\end{equation}
is derived, leaving only little room for the value of $m_0$.
 
We obtain a lower bound on the value of the $\mu$ parameter of about 500 GeV,  
although there are a few points, having inverse slepton hierarchy, 
with $\mu \sim 200$ GeV as seen in Fig.\ref{mufig}. 
The second chargino, which is almost pure Higgsino is therefore rather heavy. 
\FIGURE{
\includegraphics[width=0.45\textwidth]{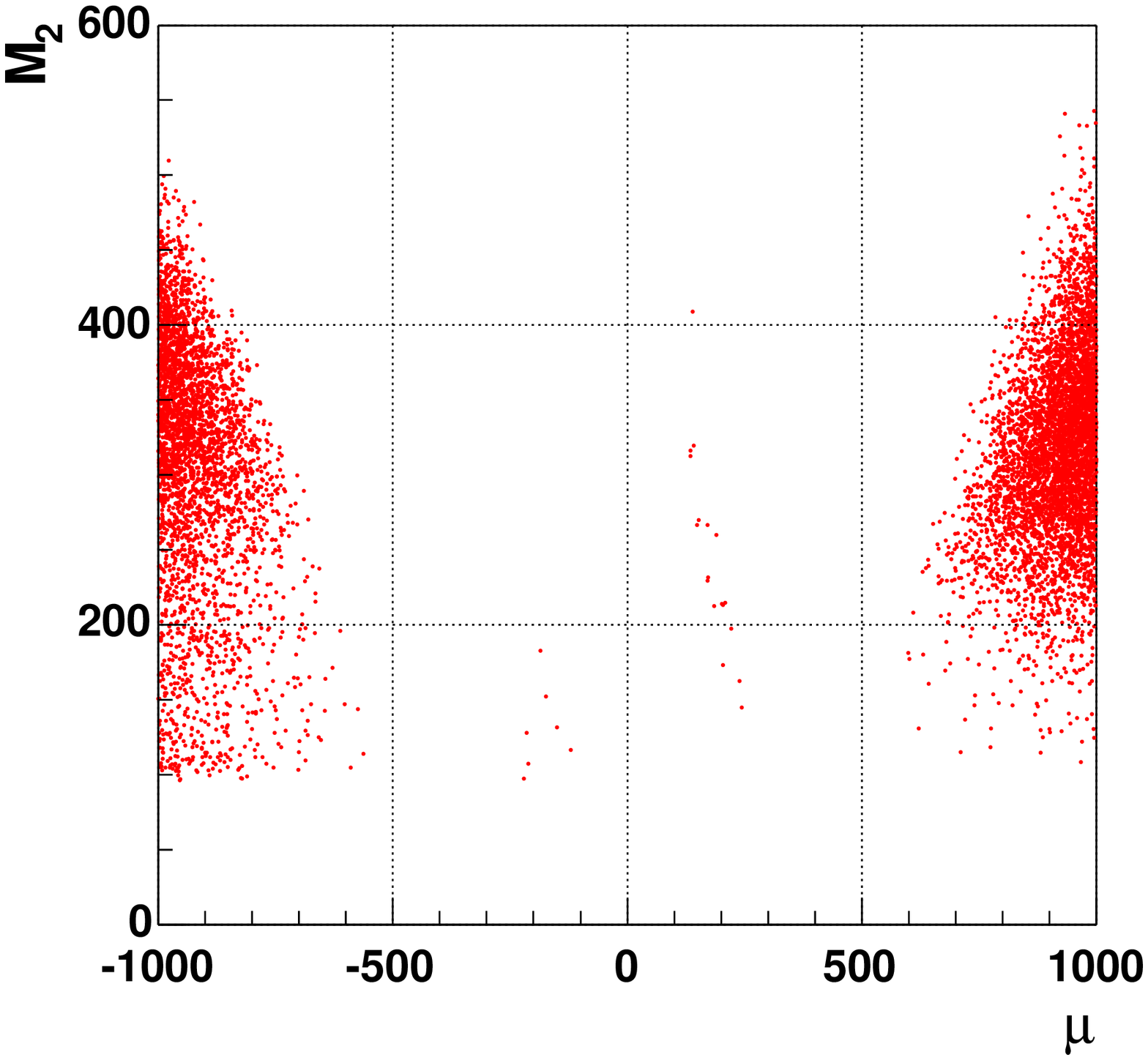}
 \caption{The allowed range for SU(2) gaugino soft breaking mass $M_2$ 
and the $\mu$-parameter for $\delta<0$. 
All points are in agreement with present bounds on LFV as well as the 
neutrino data.}
\label{mufig}}

As there is at least one low mass slepton present in the model, one could 
suspect that a large contribution to the anomalous magnetic moment 
of the muon will result. We have explicitely calculated the magnitude, 
using the formulas in Ref.\cite{magmoment}. 
The rough order of magnitude is $10 \times 10^{-10}$, 
which is too small to explain the BNL result~\cite{bnl}. 
As is well-known the contribution to $g-2$ has the same sign as 
the $\mu$-term, thereby disfavoring negative values for the $\mu$ parameter. 

An important outcome of this study is the prediction for the charged 
lepton decays $\ell_i \rightarrow \ell_j \gamma$. As seen in Fig.\ref{brfig} 
a lower bound of $10^{-9}$ for BR$(\tau \rightarrow \mu \gamma)$ 
is found. The is within reach of the future BaBar and Belle 
searches. Also, BR$(\mu \rightarrow e \gamma)$ is constrained 
to be larger than about $10^{-15}$ and therefore stands good chance of being 
observed in the near future at PSI. 

\FIGURE[t]{
  \includegraphics[width=0.45\textwidth]{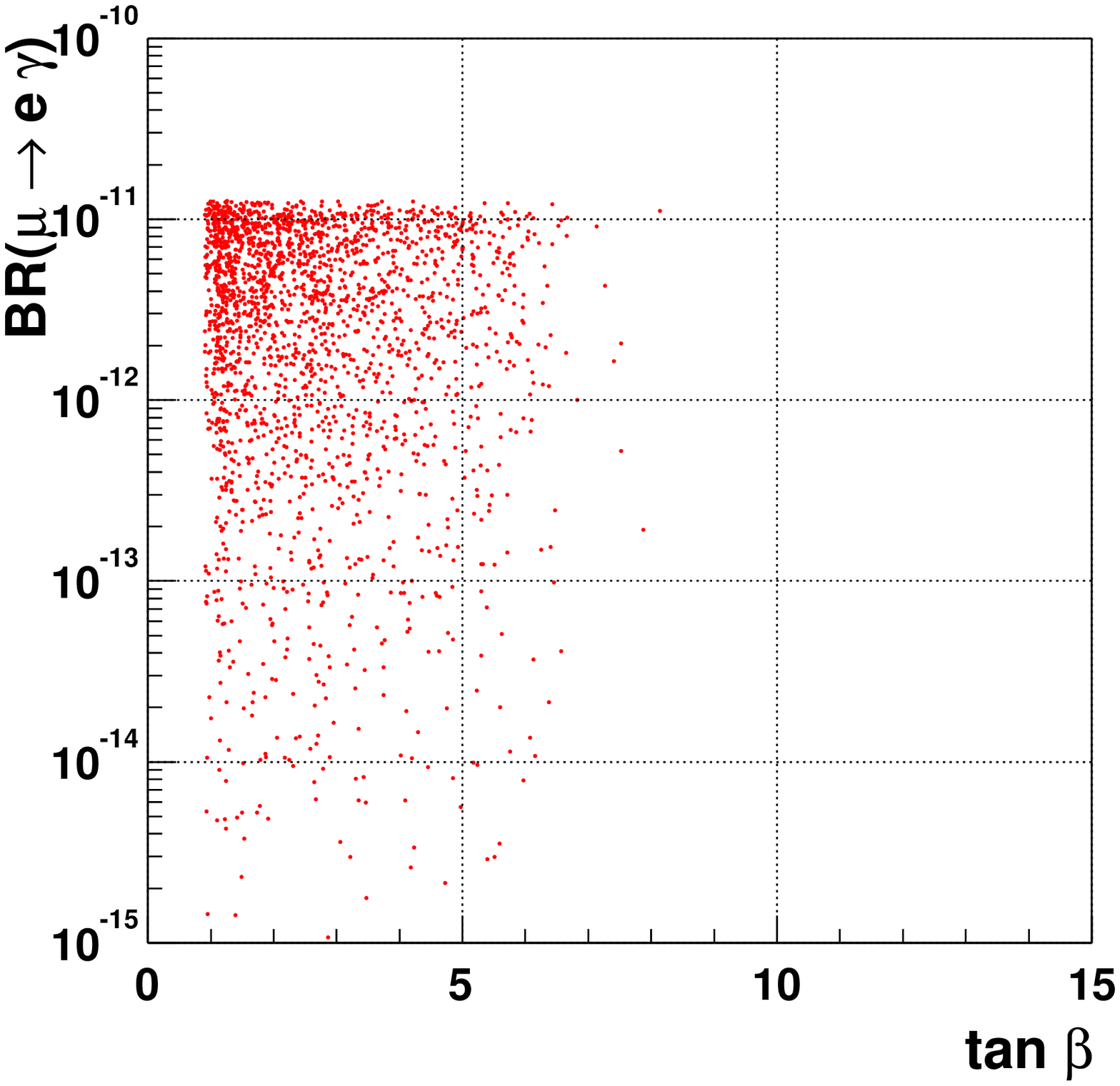}
  \includegraphics[width=0.45\textwidth]{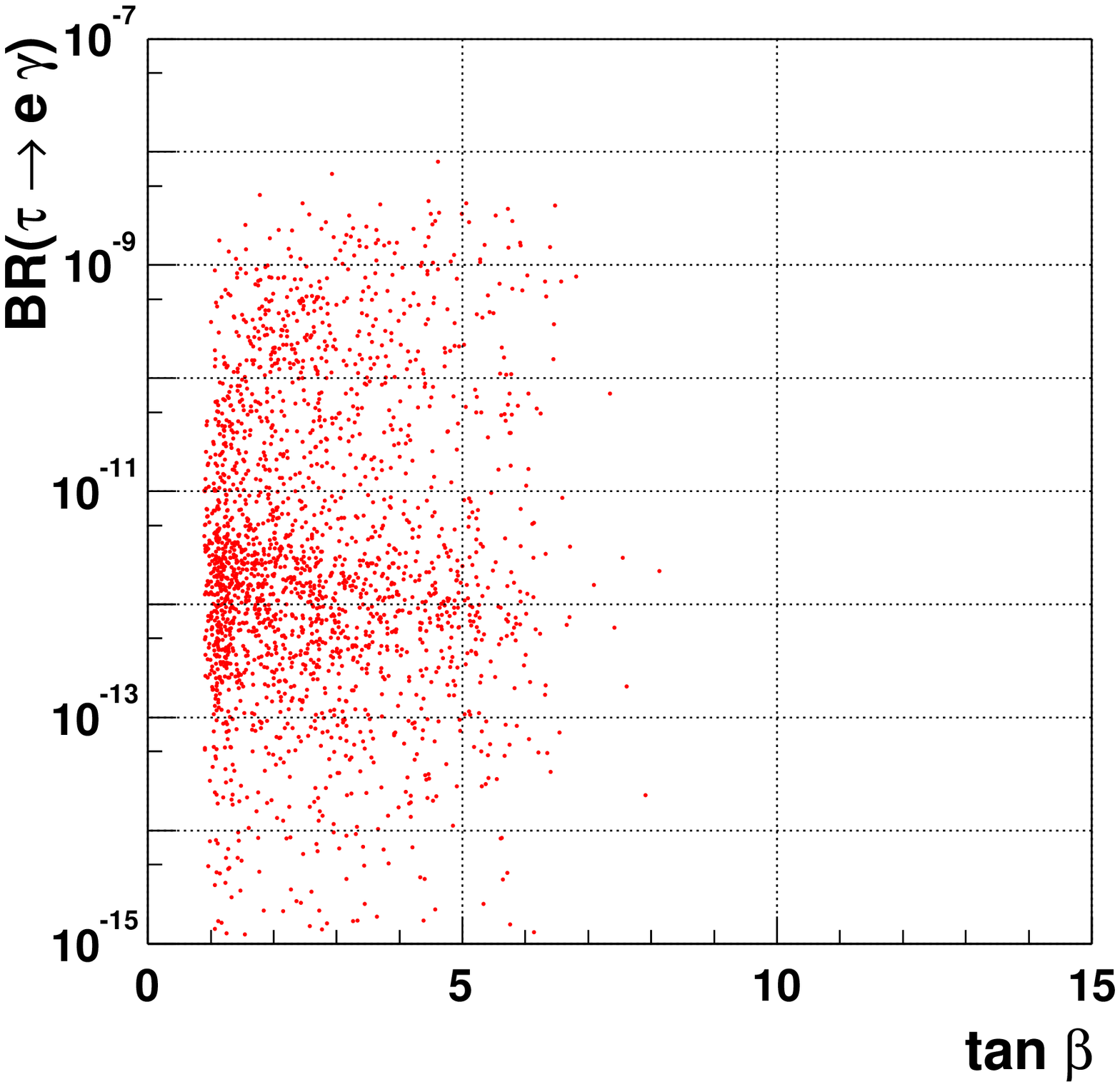}\\
  \includegraphics[width=0.45\textwidth]{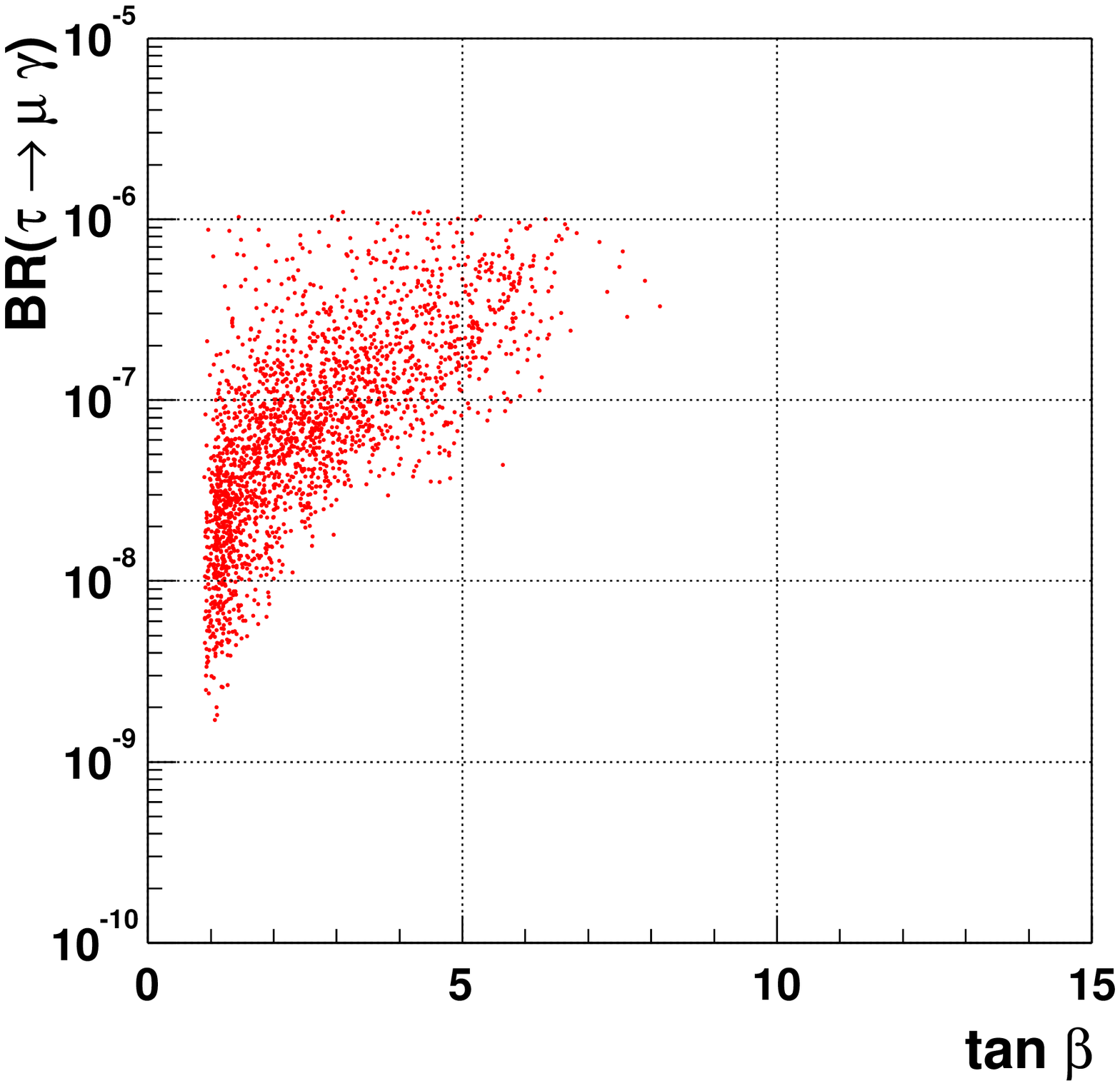}
\caption{The predictions for the branching ratios for the processes 
$\ell_i \rightarrow \ell_j \gamma$ as a function of $\tan(\beta)$. 
}
\label{brfig}
}
The value of $\tan(\beta)$, also plotted in Fig.\ref{brfig}, 
is constrained to be small. For large $\tan(\beta)$ the RGE effect 
destroys the agreement with the solar data. 
The numerical value of $\delta_\tau$ can not be much 
bigger than the solar mass scale, as seen from Eq.\ref{dmsqsol}. 
A rough estimate gives $\delta_\tau \stackrel{<}{\sim} 5 \times 10^{-4}$, 
corresponding to the bound $\tan(\beta)<10$. This agrees with the  
precise bound found in Fig.\ref{brfig}. For small values of 
$\tan(\beta)$ the threshold corrections dominate and the 
strongest constraint come from the bound on $BR(\mu \rightarrow e \gamma$). 

\section{Concluding remarks}

The supersymmetric $A_4$ model can account for the known flavor structure
and has, at leading order, the interesting features
\begin{itemize}
\item Maximal CP-violation in the leptonic sector (unless $U_{e3}=0$).
\item Maximal atmospheric mixing. 
\end{itemize}
The mixing matrices are generated by quantum corrections, starting from 
a fixed value at the high energy scale. 
The mechanism requires flavor-changing soft supersymmetry terms, 
and a close relation between the slepton and lepton sectors is obtained.  
To get sufficient suppression of LFV is a general problem in SUSY models.  
Nevertheless, as the $A_4$ model is relying on flavor violation 
in the slepton sector, the problem is more acute. 
In fact, in order to get sufficient splitting of the degenerate neutrinos,  
large mixings and large mass splittings in the slepton sector are required. 
We derive the flavor structure of the charged slepton and 
sneutrinos sectors, using neutrino data and the bounds on 
flavor-violating charged lepton decays. 
We find that the rough structure of the slepton spectrum is one almost  
degenerate pair along with one split state. Furthermore, the sleptons are 
bi-largely mixed, as are the neutrinos, 
with large mixing between $\tilde{\mu}-\tilde{\tau}$ 
and small mixing between $\tilde{e}$ and the split state.  
Therefore, the model predicts large branching 
ratio for the decay $\tau \rightarrow \mu \gamma$.  
In addition, at least one charged slepton is below 200 GeV and  
the overall neutrino mass scale is constrained to be larger 
than 0.3 eV. 

In conclusion the analysis shows that, although a functioning 
area of the SUSY parameter space exist, it is rather strongly restricted.  
If the model is realized in nature it should give rise to interesting 
experimental discoveries in the very near future. 

\acknowledgments
S.S. wishes to thank M.E.Gomez for valuable discussions. This work was 
supported by the Spanish grant BFM2002-00345, by the RTN network 
HPRN-CT-2000-00148 and by the European Science Foundation network grant N.86.

\end{document}